# A Vector-Based Representation of the Chemical Bond for the Normal Modes of Benzene

Wei Jie Huang, Alireza Azizi, Tianlv Xu, Steven R. Kirk[*], and Samantha Jenkins[*]

*Key Laboratory of Chemical Biology and Traditional Chinese Medicine Research and Key Laboratory of Resource Fine-Processing and Advanced Materials of Hunan Province of MOE, College of Chemistry and Chemical Engineering, Hunan Normal University, Changsha, Hunan 410081, China*

email: steven.kirk@cantab.net
email: samanthajsuman@gmail.com

We introduce a vector-based interpretation of the chemical bond within the quantum theory of atoms in molecules (QTAIM), the bond-path framework set $\mathbb{B} = \{p, q, r\}$, to follow variations in the 3-D morphology of all bonds for the four infra-red (IR) active normal modes of benzene. The bond-path framework set $\mathbb{B}$ comprises three unique paths $p$, $q$ and $r$ where $r$ is the familiar QTAIM bond concept of bond-path ($r$) while the two new paths $p$ and $q$ are formulated from the least and most preferred directions of electron density accumulation respectively. We find 3-D distortions including bond stretching/compression, torsion and curving. We introduce two fractional measures to quantify these variations away from linearity of the bond.

## Introduction

Two different approaches are used for the interpretation of vibrational spectroscopies using Raman Infrared(IR)[2,3] and inelastic incoherent neutron scattering of molecular structures. The first is the use of group theory with mathematical models of the forms and frequencies of the molecular vibrations including using normal mode coordinate analysis and symmetry assignments. The second is the use of empirical characteristic frequencies for chemical functional groups. We shall briefly illustrate the disadvantages of the first approach with the useful prototype of water ice, because it contains a mix of weak and strong bonding types. Previously, one of the current authors correlated the symmetry assignments of the zone-centre normal modes of the proton ordered ice VIII and the corresponding frequencies with Raman, infrared and incoherent neutron scattering spectra[3]. The comparison of the calculated frequencies with the Raman experiment spectra for the ice VIII O-H sigma bonds for the symmetric and anti-symmetric stretching modes was reasonable with only a 2% error for both of these modes. In calculations on this prototype system the errors were much larger for the bending, rotational and translational modes: up to 15%, 27% and 10% respectively. Contributions to the high frequency normal modes of vibration are dominated by the stronger O-H sigma bonds conversely; the lower frequency modes are dominated by the vibrations of the weaker hydrogen bonds and O---O bonding interactions. A reason for the much greater accuracy of the O-H sigma bond stretching mode frequencies is the lack of curving and twisting distortions of the O-H sigma bond in the spectra due to the rigidity of the O-H bond for the stretching modes. Conversely, the more flexible hydrogen bond and O---O bonding can undergo curving and twisting distortions. These types of distortions are also common across a wider range of chemical environments both in gaseous, liquid and solid states during vibrational excitations.

In this investigation we examine the four IR active normal modes of benzene and how the C-C and C-H bonds can curve and twist showing that assumptions of the rigidity of even these short bonds is an oversimplification. In other words, the geometry of the bond itself or *bond-path* is an important and overlooked aspect of the interpretation of experimental spectra. The assumption of a constantly geometrically linear bond assumes that a bond does not curve or twist and can be modelled as a spring with harmonic spring constants where anharmonic terms can be added. For bonds that can flex and curve further than the bonded inter-nuclear separation, this is clearly no longer a valid model, even when anharmonic terms are included. Additional oversimplifications are that torsion motions occur as rigid atom-centred distortions that do not involve any twisting of the bond.

Recently, a methodology[4] was created to gain new insights into the subtle details of the four infrared (IR) active normal modes of vibration of benzene[5–10]. This approach based on the quantum theory of atoms in molecules QTAIM[11] and the stress tensor trajectories $\mathbb{T}_\sigma$ in the stress tensor eigenvector projection space $\mathbb{U}_\sigma$[12,13]. We used the new methodology to explain the relative differences in the intensities of the IR active modes associated with changes in the dipole moment[6–8,14,15] where larger dipole moments led to greater intensities of the normal mode of vibrations. This work was intended to complement previous extensive studies of the vibrational modes of benzene using semi-empirical and *ab initio* methods[16–18] and theoretical treatments on the normal coordinate treatment of benzene[5,19,20], see **Scheme 1**.

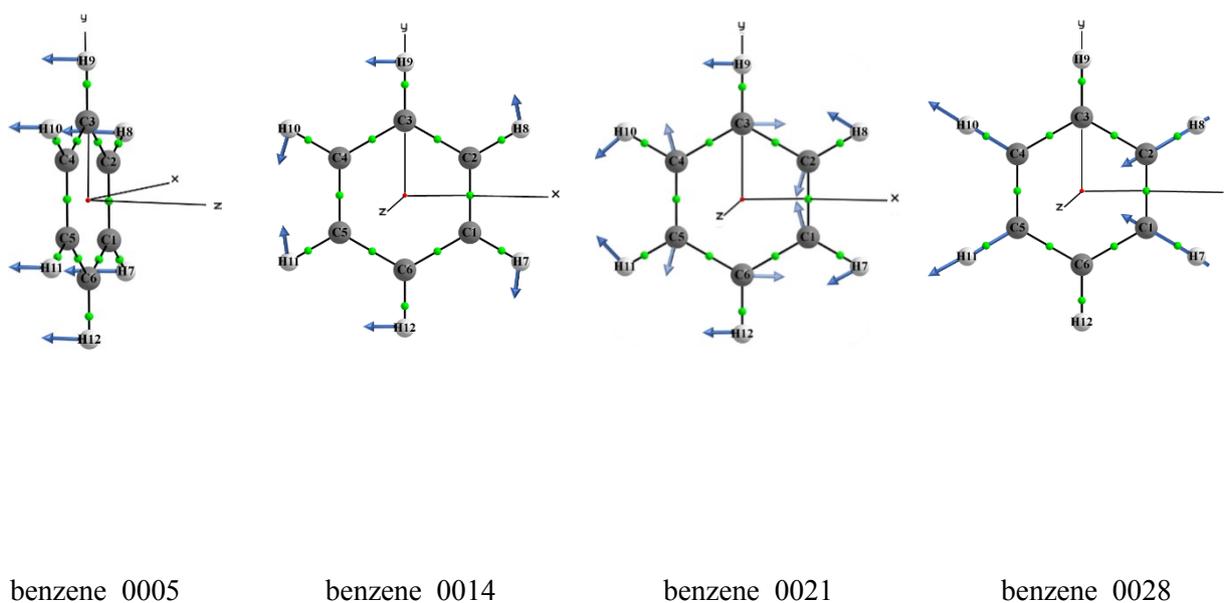

benzene_0005　　　　benzene_0014　　　　benzene_0021　　　　benzene_0028

**Scheme 1**. The conventional rendering of four normal modes of benzene are superimposed onto the molecular graphs of the relaxed structure of benzene. The four infrared (IR) active normal modes of benzene are ordered according to their increasing frequency, corresponding to 721.568 cm$^{-1}$, 1097.691 cm$^{-1}$, 1573.927 cm$^{-1}$ and 3298.320 cm$^{-1}$ respectively, the arrows are not drawn to scale. The corresponding intensities (10$^{-7}$ cm$^2$ mol.$^{-1}$s$^{-1}$ ln) are 136.47, 5.78, 10.92 and 24.86 respectively.

In this investigation we use a new vector-based interpretation of the chemical bond that considers the entire bond as opposed to only properties derived at the *BCP*, as was the case for our earlier investigations, of the normal modes of vibration of the IR active modes of benzene. The reason to consider the entire bond-path is to investigate all of the 3-D variations that will be non-linear in the C-H and C-C bond topology including bond curving and twisting for each of the four IR active normal modes. This will show the reasons for the insufficiency of current mathematical models used for analyzing experimental vibrational spectra that rely on the *ad-hoc* addition of anharmonic terms, whilst assuming the presence of linear bonds.

## 2. Theory and Methods

*2.1 The QTAIM critical point properties; the bond critical point (BCP) ellipticity ε and the molecular graph*

We use QTAIM and the stress tensor analysis that utilizes higher derivatives of $\rho(\mathbf{r}_b)$ in effect, acting as a

'magnifying lens' on the $\rho(\mathbf{r_b})$ derived properties of the wave-function. We will use QTAIM[11] to identify critical points in the total electronic charge density distribution $\rho(\mathbf{r})$ by analyzing the gradient vector field $\nabla\rho(\mathbf{r})$. These critical points can further be divided into four types of topologically stable critical points according to the set of ordered eigenvalues $\lambda_1 < \lambda_2 < \lambda_3$, with corresponding eigenvectors $\underline{\mathbf{e}}_1$, $\underline{\mathbf{e}}_2$, $\underline{\mathbf{e}}_3$ of the Hessian matrix. The Hessian of the total electronic charge density $\rho(\mathbf{r})$ is defined as the matrix of partial second derivatives with respect to the spatial coordinates. The eigenvector $\underline{\mathbf{e}}_3$ indicates the direction of the bond-path at the *BCP*. The most and least preferred directions of electron accumulation are $\underline{\mathbf{e}}_2$ and $\underline{\mathbf{e}}_1$, respectively[21–23]. The ellipticity, ε provides the relative accumulation of $\rho(\mathbf{r_b})$ in the two directions perpendicular to the bond-path at a *BCP*, defined as $\varepsilon = |\lambda_1|/|\lambda_2| - 1$ where $\lambda_1$ and $\lambda_2$ are negative eigenvalues of the corresponding eigenvectors $\underline{\mathbf{e}}_1$ and $\underline{\mathbf{e}}_2$ respectively.

The four types of critical points are labeled using the notation (*R*, ω) where *R* is the rank of the Hessian matrix, the number of distinct non-zero eigenvalues and ω is the signature (the algebraic sum of the signs of the eigenvalues); the (3, -3) [nuclear critical point (*NCP*), a local maximum generally corresponding to a nuclear location], (3, -1) and (3, 1) [saddle points, called bond critical points (*BCP*) and ring critical points (*RCP*), respectively] and (3, 3) [the cage critical points (*CCP*)]. In the limit that the forces on the nuclei become vanishingly small, an atomic interaction line[24] becomes a bond-path, although not necessarily a chemical bond[25]. The complete set of critical points together with the bond-paths of a molecule or cluster is referred to as the molecular graph.

*2.2 The QTAIM bond-path properties; BPL, bond-path curvature, eigenvector-following path lengths $\mathbb{H}$, $\mathbb{H}^*$ and the bond-path framework set $\mathbb{B}$*

The bond-path length (BPL) is defined as the length of the path traced out by the $\underline{\mathbf{e}}_3$ eigenvector of the Hessian of the total charge density $\rho(\mathbf{r})$, passing through the *BCP*, along which $\rho(\mathbf{r})$ is locally maximal with respect to any neighboring paths. The bond-path curvature separating two bonded nuclei is defined as the dimensionless ratio:

(BPL - GBL)/GBL                                                                                                                              **(1)**

Where BPL defined to be the bond-path length associated and GBL is the inter-nuclear separation. The BPL often exceeds the GBL particularly in for weak or strained bonds and unusual bonding environments[26]. Earlier, one of the current authors hypothesized that a bond-path may possess 1-D, 2-D or a 3-D morphology[27], with 2-D or a 3-D bond-paths associated with a *BCP* with ellipticity ε > 0, being due to the differing degrees of charge density accumulation, associated with the $\lambda_2$ and $\lambda_1$ eigenvalues respectively. Bond-paths possessing zero and non-zero values of the bond-path curvature defined by equation **(1)** can be considered to possess 1-D and 2-D topologies respectively. We start by choosing the length traced out in 3-D by the path swept by the tips of the scaled $\underline{\mathbf{e}}_2$ eigenvectors of the $\lambda_2$ eigenvalue, the scaling factor being chosen as the ellipticity $\varepsilon$.

With $n$ scaled eigenvector $\underline{e}_2$ tip path points $q_i = r_i + \varepsilon_i\underline{e}_{2,i}$ on the path $q$ where $\varepsilon_i$ = ellipticity at the $i^{th}$ bond-path point $r_i$ on the bond-path $r$, see equation (**2a**). It should be noted that the bond-path is associated with the $\lambda_3$ eigenvalues of the $\underline{e}_3$ eigenvector does not take into account differences in the $\lambda_1$ and $\lambda_2$ eigenvalues of the $\underline{e}_1$ and $\underline{e}_2$ eigenvectors. Analogously, equation (**2b**), is used for the $\underline{e}_1$ tip path points we have $p_i = r_i + \varepsilon_i\underline{e}_{1,i}$ on the path $p$ where $\varepsilon_i$ = ellipticity at the $i^{th}$ bond-path point $r_i$ on the bond-path $r$.

We will refer to the new QTAIM interpretation of the chemical bond as the *bond-path framework set* that will be denoted by $\mathbb{B}$, where $\mathbb{B} = \{p, q, r\}$. This effectively means that in the most general case a bond is comprised of three 'linkages'; $p$, $q$ and $r$ associated with the $\underline{e}_1$, $\underline{e}_2$ and $\underline{e}_3$ eigenvectors respectively.

From this we shall define eigenvector-following path length $\mathbb{H}^*$ and $\mathbb{H}$, see **Scheme 2**:

$$\mathbb{H}^* = \sum_{i=1}^{n-1}|p_{i+1} - p_i| \qquad (2a)$$
$$\mathbb{H} = \sum_{i=1}^{n-1}|q_{i+1} - q_i| \qquad (2b)$$

The *eigenvector-following path* length $\mathbb{H}^*$ or $\mathbb{H}$ refers to the fact that the tips of the scaled $\underline{e}_1$ or $\underline{e}_2$ eigenvectors will sweep out along the extent of the bond-path, defined by the $\underline{e}_3$ eigenvector, between the two bonded nuclei that the bond-path connects.

From the form of $p_i = r_i + \varepsilon_i\underline{e}_{1,i}$ and $q_i = r_i + \varepsilon_i\underline{e}_{2,i}$ we see for shared-shell *BCP*s, that in the limit of the ellipticity $\varepsilon \approx 0$ *for all* steps $i$ along the bond-path i.e. corresponding to single bonds, we then have $p_i = q_i = r_i$ and therefore the value of the lengths $\mathbb{H}^*$ and $\mathbb{H}$ attain their lowest limit; the bond-path length ($r$) BPL. Conversely, higher values of the ellipticity $\varepsilon$, for instance, corresponding to double bonds will always result in values of $\mathbb{H}^*$ and $\mathbb{H} >$ BPL.

The bond-path framework set $\mathbb{B} = \{p, q, r\}$ considers the bond-path to comprise the *unique* paths, $p$, $q$ and $r$, swept out by the $\underline{e}_1$, $\underline{e}_2$ and $\underline{e}_3$, eigenvectors that form the eigenvector-following path lengths $\mathbb{H}^*$, $\mathbb{H}$ and BPL respectively. The paths $p$ and $q$ are unique even when the lengths of $\mathbb{H}^*$ and $\mathbb{H}$ are the same or very similar because $p$ and $q$ traverse different regions of space.

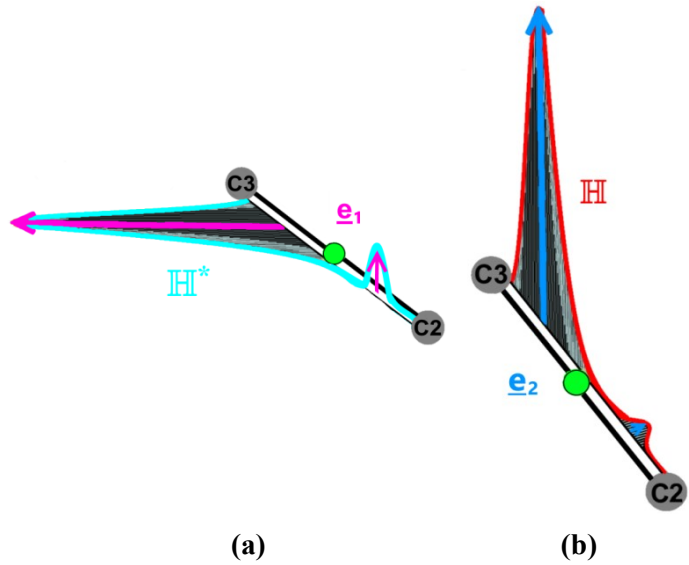

(a) (b)

**Scheme 2.** The pale blue line in sub-figure **(a)** represents the path, referred to as the eigenvector-following path length $\mathbb{H}^*$, swept out by the tips of the scaled $\underline{e}_1$ eigenvectors, shown in magenta, and defined by equation **(2a)**. The red path in sub-figure **(b)** corresponds to the eigenvector-following path length $\mathbb{H}$, constructed from the tips of the scaled $\underline{e}_2$ eigenvectors, shown in mid-blue and is defined by equation **(2b)**. The pale and mid-blue arrows representing the $\underline{e}_1$ and $\underline{e}_2$ eigenvectors are scaled by the ellipticity ε respectively, where the vertical scales are exaggerated for visualization purposes. The green sphere indicates the position of a given *BCP*. Details of how to implement the calculation of the eigenvector-following path lengths $\mathbb{H}^*$ and $\mathbb{H}$ are provided in the **Supplementary Materials S7**.

Analogous to the bond-path curvature, see equation **(1)**, we may define dimensionless, *fractional* versions of the eigenvector-following path length $\mathbb{H}$ where several forms are possible and not limited to the following:

$\mathbb{H}_f = (\mathbb{H} - \text{BPL})/\text{BPL}$            **(3a)**

$\mathbb{H}_{f\theta min} = (\mathbb{H} - \mathbb{H}_{\theta min})/\mathbb{H}_{\theta min}$            **(3b)**

Where $\mathbb{H}_{\theta min}$ is defined as the length swept out by the scaled $\underline{e}_2$ eigenvectors using the using the value of the torsion θ at the energy minimum, with similar expressions for $\mathbb{H}^*_f$ and $\mathbb{H}^*_{f\theta min}$ can be derived using the $\underline{e}_1$ eigenvectors.

The form of $\mathbb{H}_f$ defined by equation **(3a)** is the closest to the spirit of the bond-path curvature, equation **(1)**. The bond-path framework set $\mathbb{B}$ is constructed from the orthogonal triad of the $\underline{e}_1$, $\underline{e}_2$ and $\underline{e}_3$ eigenvectors and therefore will be suitable for capturing the 3-D motion of any of the normal modes, in this case of the benzene molecule. We can compare the effect on the {$\underline{e}_1$, $\underline{e}_2$, $\underline{e}_3$} bond-path frameworks of the C-C *BCP*s and C-H *BCP*s for the four IR active normal modes of benzene. This will be undertaken in terms of the bond torsion and stretching/compression of the bond-path framework set $\mathbb{B}$; $\mathbb{H}^*$, $\mathbb{H}$ and BPL respectively as well as the corresponding fractional measures ($\mathbb{H}^*_f, \mathbb{H}_f$), ($\mathbb{H}^*_{f\theta min}, \mathbb{H}_{f\theta min}$) and bond-path curvature. For a non-torsional distortion such as a pure bond stretching/compression motion there is no change in orientation the $\underline{\mathbf{e}}_1$ and $\underline{\mathbf{e}}_2$ eigenvectors of the $\{\underline{\mathbf{e}}_1, \underline{\mathbf{e}}_2, \underline{\mathbf{e}}_3\}$ bond-path framework for all values of the scaling factor $\varepsilon_i$. A consequence of this is that the $\mathbb{H}^*$ and $\mathbb{H}$ a linear variation of $\mathbb{H}^*$ or $\mathbb{H}$ with amplitude implies that there is no torsion θ present in the bond-path distortion due to the construction of with the $\mathbb{H}^*$ or $\mathbb{H}$ from the $\underline{\mathbf{e}}_1$ and $\underline{\mathbf{e}}_2$ eigenvectors respectively. Conversely, this means that that a *non-linear* variation of $\mathbb{H}^*$ or $\mathbb{H}$ with amplitude implies some degree of bond-path torsion θ is present in the bond-path distortion.

In this investigation we shall examine the eigenvector-following path lengths $\mathbb{H}$ and $\mathbb{H}^*$ for all of the *BCP*s of the benzene bond-paths, all of which correspond to shared-shell *BCP*s. As was mentioned previously, the ellipticity $\varepsilon_i$ values along the bond-paths (*r*) associated with shared-shell *BCP*s are chemically meaningful. As a consequence, we suggest that for a linear variation of $\mathbb{H}$ or $\mathbb{H}^*$ with amplitude that longer/shorter $\mathbb{H}$ or $\mathbb{H}^*$ lengths compared with a given bond-path length (*r*) are due to the presence of higher/lower ellipticity ε values. Higher/lower ellipticity ε values indicate a greater/lesser participation of a C-C *BCP* or C-H *BCP* during the vibrational mode as the bond-paths are compressed/stretched due to shifts in the $\rho(\mathbf{r})$ distribution along an entire bond-path (*r*). For a *non-linear* variation of $\mathbb{H}^*$ or $\mathbb{H}$ with amplitude we have the added complexity of bond-path torsion distortions that occur with higher degrees of non-linearity in the variation of $\mathbb{H}^*$ or $\mathbb{H}$ with amplitude. Larger differences between $\mathbb{H}^*$ or $\mathbb{H}$ will occur as a result of a greater degree of bond-path torsion. In addition, bond-path compression and stretching distortions will also be present for a non-linear variation of $\mathbb{H}^*$ or $\mathbb{H}$ with amplitude.

## 3. Computational Methods

The geometry optimization and frequency calculation were carried out using the BHandHLYP DFT functional and 6-31G(d,p) basis set in G09 v.E.01[28]. The SCF convergence criteria were set to tighter values than the defaults, specifically to both $< 10^{-10}$ RMS change in the density matrix and $< 10^{-8}$ maximum change in the density matrix. Single-point calculations, corresponding to snapshots throughout one full cycle of each of the calculated vibrational modes, were carried out using the same functional, basis set and convergence criteria. The intermediate 'snapshot' structures were generated by an external program using the geometry-optimized structure together with the displacement directions calculated for each mode in the frequency calculation. The wave-functions produced by the snapshot single-point calculation were then analyzed using AIMAll[29] and the resulting molecular graphs analyzed using two in-house codes, to calculate the eigenvector-following path lengths $\mathbb{H}$ and $\mathbb{H}^*$, see **Supplementary Materials S4**.

# 4. Results and Discussions

Inspection of the variation of the C-C *BCP* bond-path lengths (BPLs) with amplitude of the four normal modes; mode 0005, mode 0014, mode 0021 and mode 0028, demonstrates the extent of the participation of bond-paths of the C-C *BCP*s for each of the normal modes, see **Figure 1(a-d)** respectively. It can be seen that for all of the C-C *BCP*s of all four normal modes that the eigenvector-following path lengths $\mathbb{H}$ exceed those of the corresponding BPLs, compare the red and black plot lines in **Figure 1** respectively. The explanation for this behavior is that all of the C-C *BCP*s have sufficiently high values for the ellipticity ε, see also equation **(2a)** where $q_i = r_i + \varepsilon_i \underline{e}_{2,i}$ and $r_i$ denotes the BPL. This is simply because from the form of the $q_i$, larger ellipticity $\varepsilon_i$ values result in the length of the $\mathbb{H}$ values to significantly exceed the BPL ($r_i$).

The variations of all of the C-C *BCP*s bond-path lengths (BPLs) with amplitude are linear for all for normal modes of benzene, see the black plots in **Figure 1**. For the lowest frequency IR active mode; mode 0005 we see that, as expected, the BPL variations with the amplitude are absent for the C-C *BCP*s, see **Figure 1(a)**. For mode 0014 and mode 0021 the difference between the BPL variations with the amplitude and the corresponding $\mathbb{H}$ values, see the red plots, for a given *BCP* is a linear translation, see **Figure 1(b)** and **Figure 1(c)** respectively. The consequence is that for mode 0014 and mode 0021 the motion of the bond-paths of C-C *BCP*s of these benzene IR active modes is restricted to bond-path stretching/compression motions, see **Figure 1(b)** and **Figure 1(c)** respectively. Conversely, we find this is not the case for mode 0005 and mode 0028, see **Figure 1(a)** and **Figure 1(d)** respectively. We suggest that for mode 0005 and mode 0028 tracking the variation of the BPL along the bond-path (directed along $\underline{e}_3$) with amplitude captures the bond-path stretching and compressing motions whereas $\mathbb{H}$ captures the torsional motions. The motion of the of the C-C *BCP* bond-paths for the 0005 mode and the 0028 modes is a mix of torsional motion in addition to stretching/compression motions, although the distortion of the mode 0005 mode C-C *BCP* bond-paths is very low, **Figure 1(a)** and **Figure 1(d)** respectively. For the C-C *BCP* bond-paths there are small differences visible between $\mathbb{H}$ and $\mathbb{H}^*$ plots for mode 0028, compare **Figure 1(d)** with **Figure S1(d)** of the **Supplementary Materials S1**.

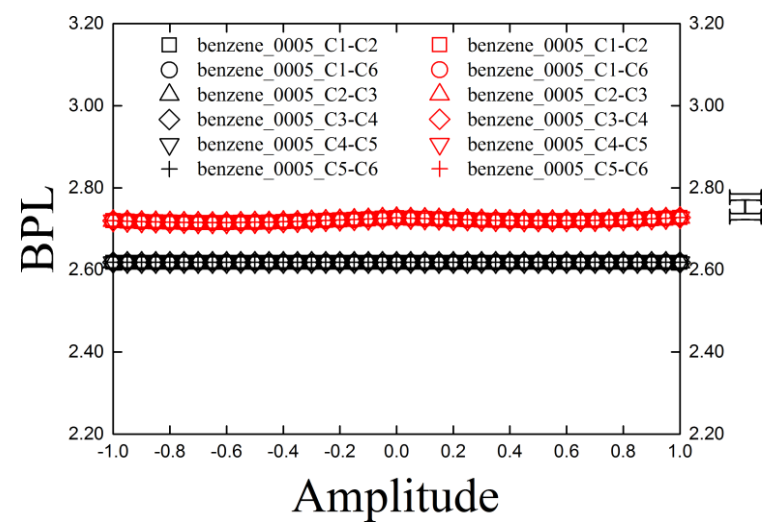 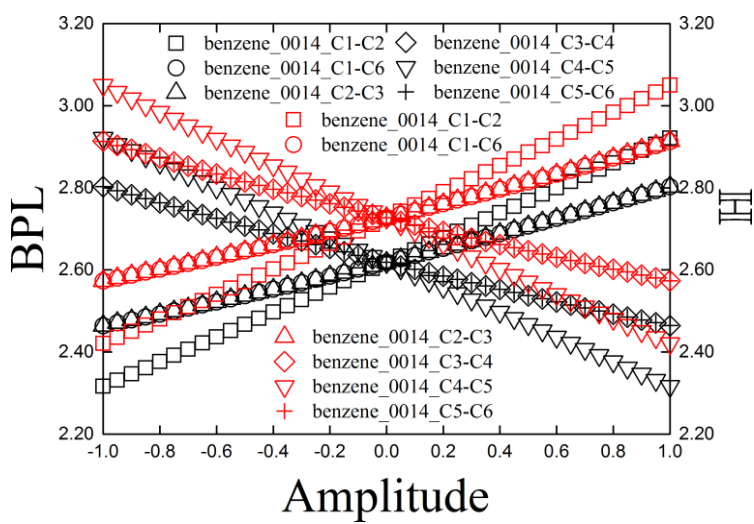

(a) (b)

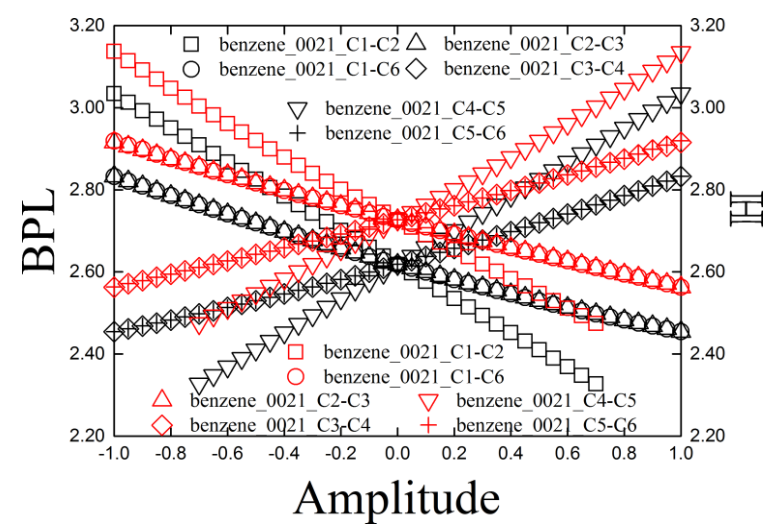 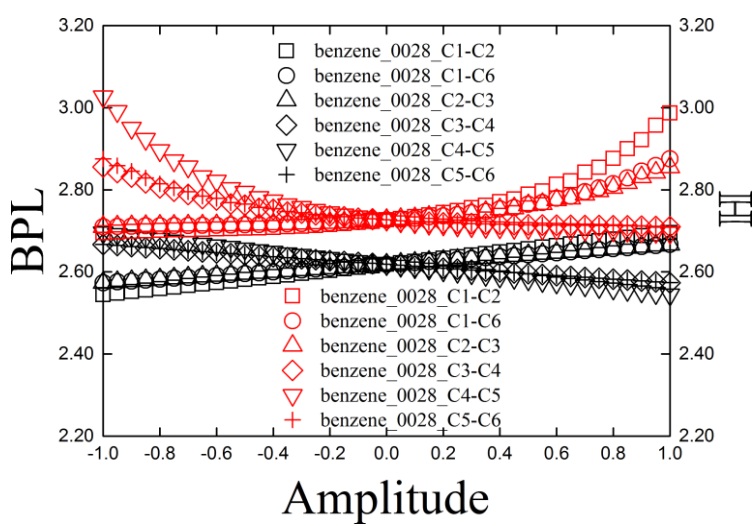

(c) (d)

**Figure 1.** The variation of the bond-path lengths (BPL) and the eigenvector-following path length $\mathbb{H}$ of the C-C *BCPs* with amplitude for the four benzene infrared (IR) active modes are shown in black and red respectively in sub-figures **(a)-(d).** The corresponding variations of the eigenvector-following path length $\mathbb{H}^*$ are provided in the **Supplementary Materials S1**.

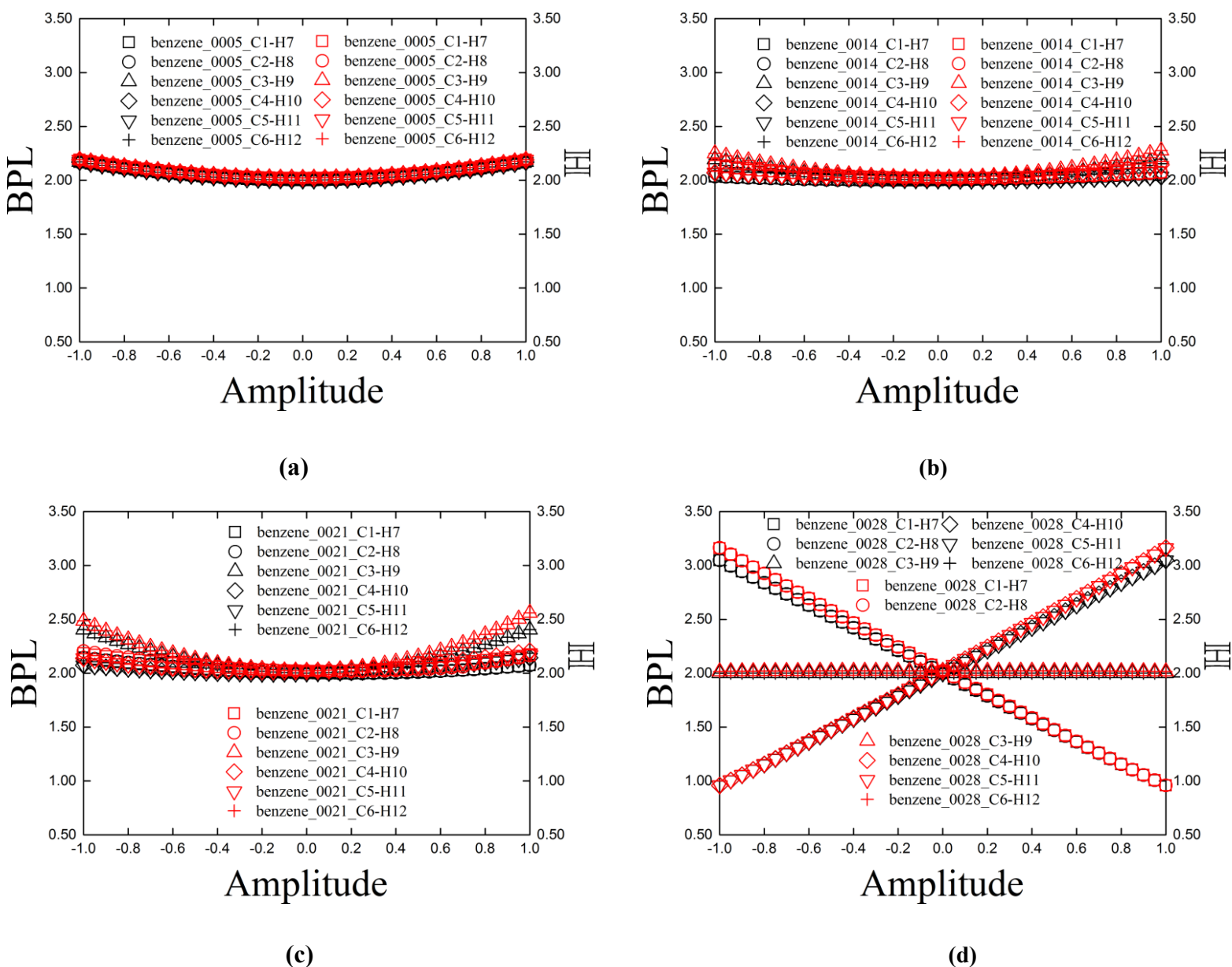

**(a)** **(b)** **(c)** **(d)**

**Figure 2.** The variation of the bond-path lengths (BPL) of the C-H *BCPs* with amplitude for the four IR active modes of benzene are shown in sub-figures **(a)-(d)** respectively, see the caption of **Figure 1** for further details. The corresponding variations of the eigenvector-following path length $\mathbb{H}^*$ are provided in the **Supplementary Materials S2**.

The BPL variations with the amplitude for the normal modes of vibration for the C-H *BCP*s capture the increasingly strong relative motion of the C-H bond-path from the lowest to the highest frequency, see the black plots in **Figure 2(a-d)** respectively. Differences between $\mathbb{H}$ and $\mathbb{H}^*$ are apparent for the C-H bond-paths of mode 0014 and mode 0021, compare the red plots in **Figure 2(b-c)** with **Figure S2(b-c)** of the **Supplementary Materials S2**. Differences between $\mathbb{H}$ and $\mathbb{H}^*$ are due to non-zero bond-path curvatures, see the theory section 2.2 for more explanation.

Consistency is found with our recent publication with the stress tensor trajectories and these four IR active normal modes; we again find that for mode 0014 and mode 0028 the C-C *BCP*s excess BPL distortions indicate an equal or at least significant contribution to the normal mode of vibration. Conventional normal mode analysis determines that both modes only involve the C-H bonds and again we find this is clearly an over simplification as is apparent for the non-zero variation in the BPL for the C-C *BCP*s with amplitude, see **Figure 1(b)** and **Figure 1(d)** respectively.

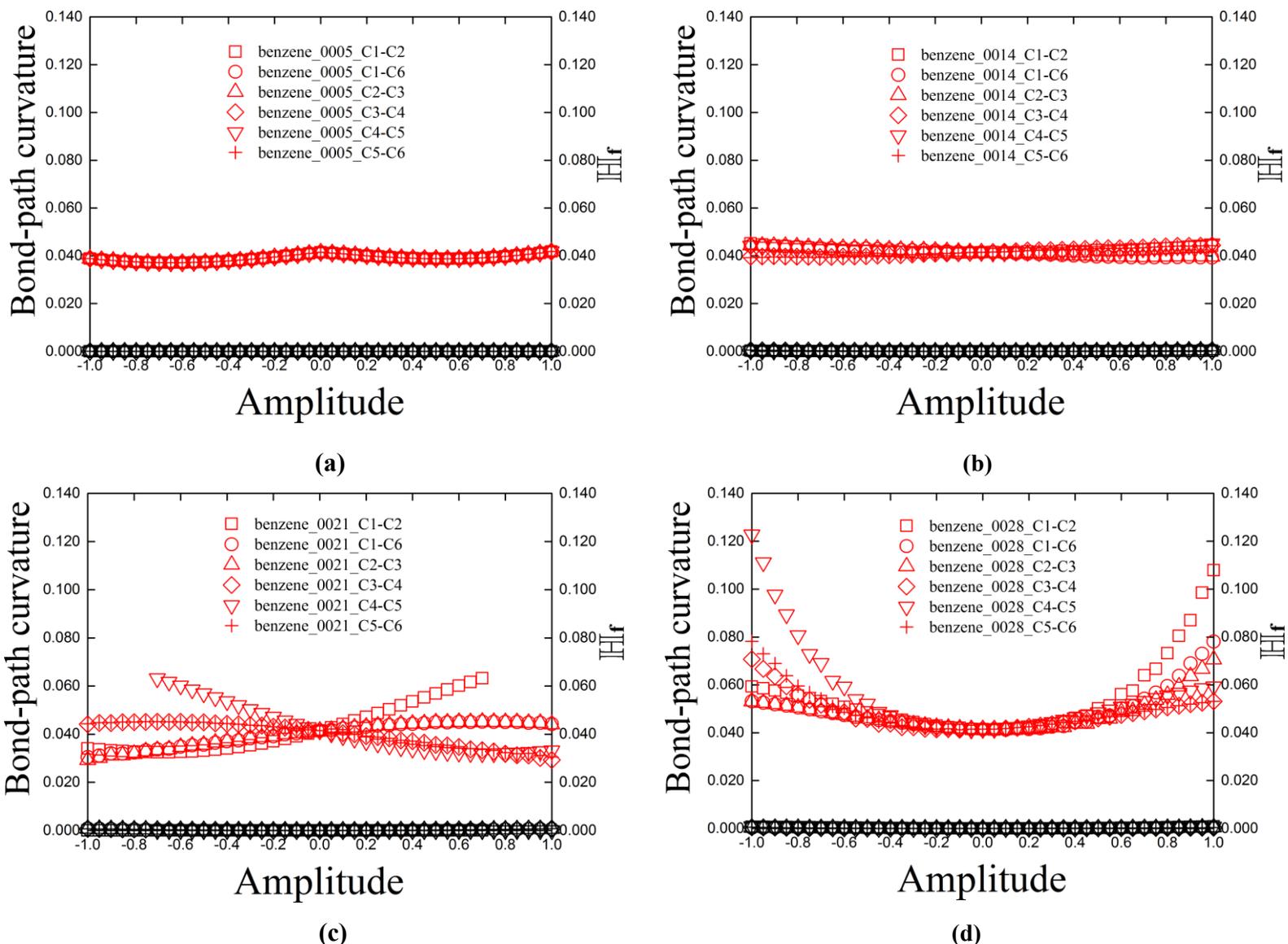

(a)

(b)

(c)

(d)

**Figure 3.** The variation of the bond-path curvature (BPL - GBL)/GBL, see equation **(1)** and the eigenvector-following path length $\mathbb{H}_f$ of the C-C *BCPs* with amplitude for the for the four IR active modes of benzene are shown in sub-figures **(a)** and **(d)** respectively. The corresponding variations the fractional eigenvector-following path length $\mathbb{H}^*_f$ are provided in the **Supplementary Materials S3**.

The presence of non-zero bond-path curvatures associated with the C-C *BCP*s is not apparent, however, the fractional eigenvector-following path length $\mathbb{H}_f$ do display significant values, see **Figure 3**. In addition, the previously noted differences between $\mathbb{H}$ and $\mathbb{H}^*$ plots for mode 0028 are more apparent for the fractional variants $\mathbb{H}_f$ and $\mathbb{H}^*_f$ plots for mode 0028, as well as differences for mode 14 and mode 21 compare **Figure 3(d)** with **Figure S2(d)** of the **Supplementary Materials S3**. We also observe a non-linear variation of the

$\mathbb{H}_f$ for mode 0021 that was not apparent for the $\mathbb{H}$ values, due to the fractional $\mathbb{H}_f$ magnifying small variations in the data.

For the $\mathbb{H}$ values of mode 0005 we see very small amplitude variations, visible as a slight peak for the equilibrium configuration, for the eigenvector-following path length $\mathbb{H}$ that is magnified for the fractional eigenvector-following path length $\mathbb{H}_f$, see **Figure 1(a)** and **Figure 3(a)** respectively.

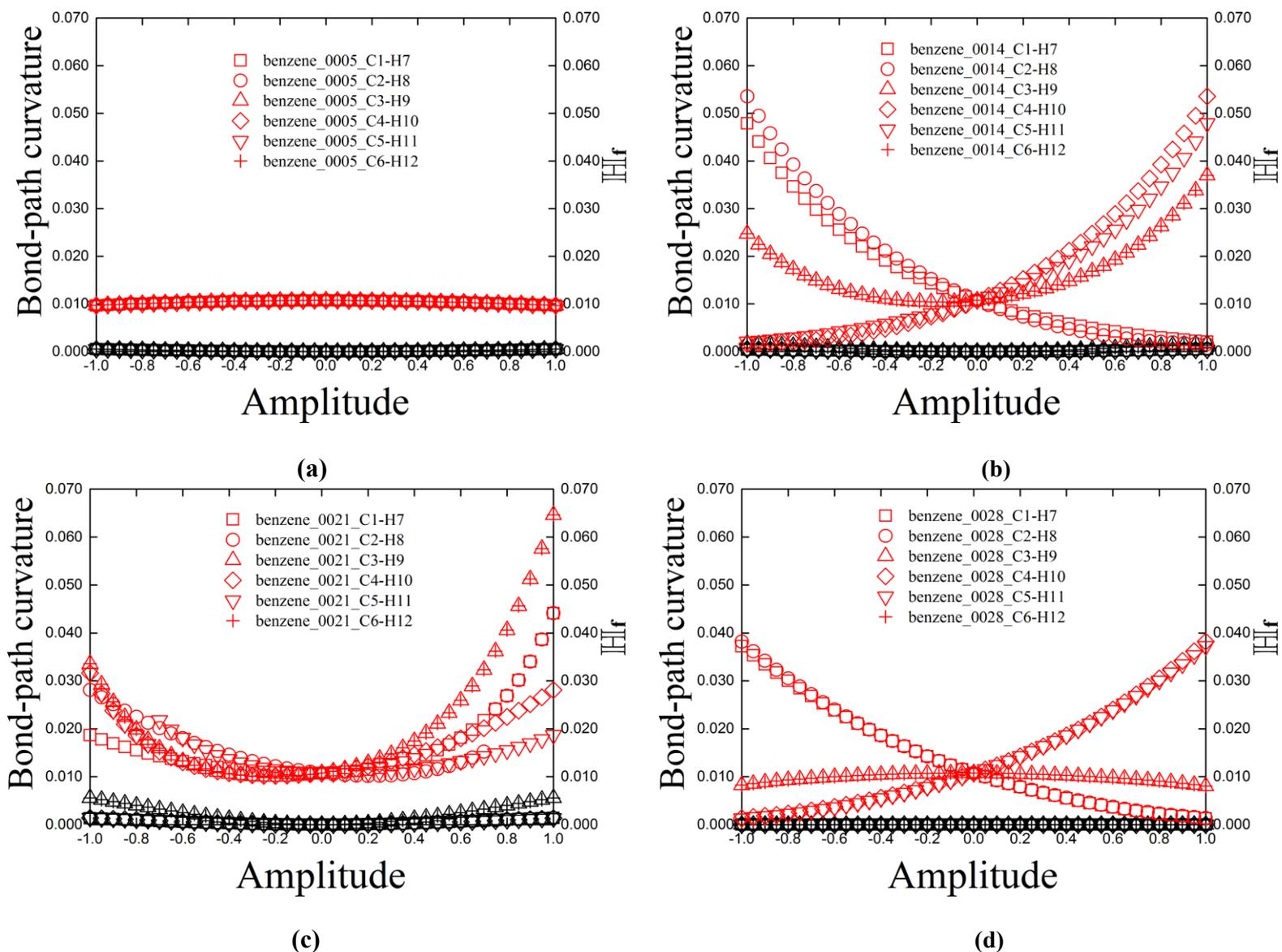

**Figure 4.** The variation of the bond-path curvature and the eigenvector-following path length $\mathbb{H}_f$ of the C-H *BCPs* with amplitude for the four IR active modes of benzene are shown in the left and right hand panels of sub-figures **(a)** and **(d)** respectively, see the caption of **Figure 3** for further details. The corresponding variations the fractional eigenvector-following path length $\mathbb{H}^*_f$ are provided in the **Supplementary Materials S4**.

For all four modes; the bond-paths associated with the C-H *BCPs* possess values of $\mathbb{H}$ that are almost degenerate with the BPLs because all of the C-H BCPs possess low ellipticity values ε for the benzene C-H

BCPs, seen from the expression for the from $q_i = r_i + \varepsilon_i \underline{e}_{2,i}$ that determines the eigenvector-following path length $\mathbb{H}$, see **Figure 2**. Of the four modes only the C-H BCPs of the highest frequency mode 0028 have linear variations of the BPL with respect to the amplitude. The reason is that for mode 0028 the C-H bond-paths are distorted linearly i.e. bond stretching/compression, without a significant degree of bond-path torsion resulting from the zero bond-path curvature throughout the duration of the normal mode. The non-zero variations in the bond-path curvature of mode 0021 were apparent from the molecular graphs of the maximum ± amplitudes of the normal modes of vibration of our earlier work[4] therefore we provide these plots for reference in the **Supplementary Materials S8.**

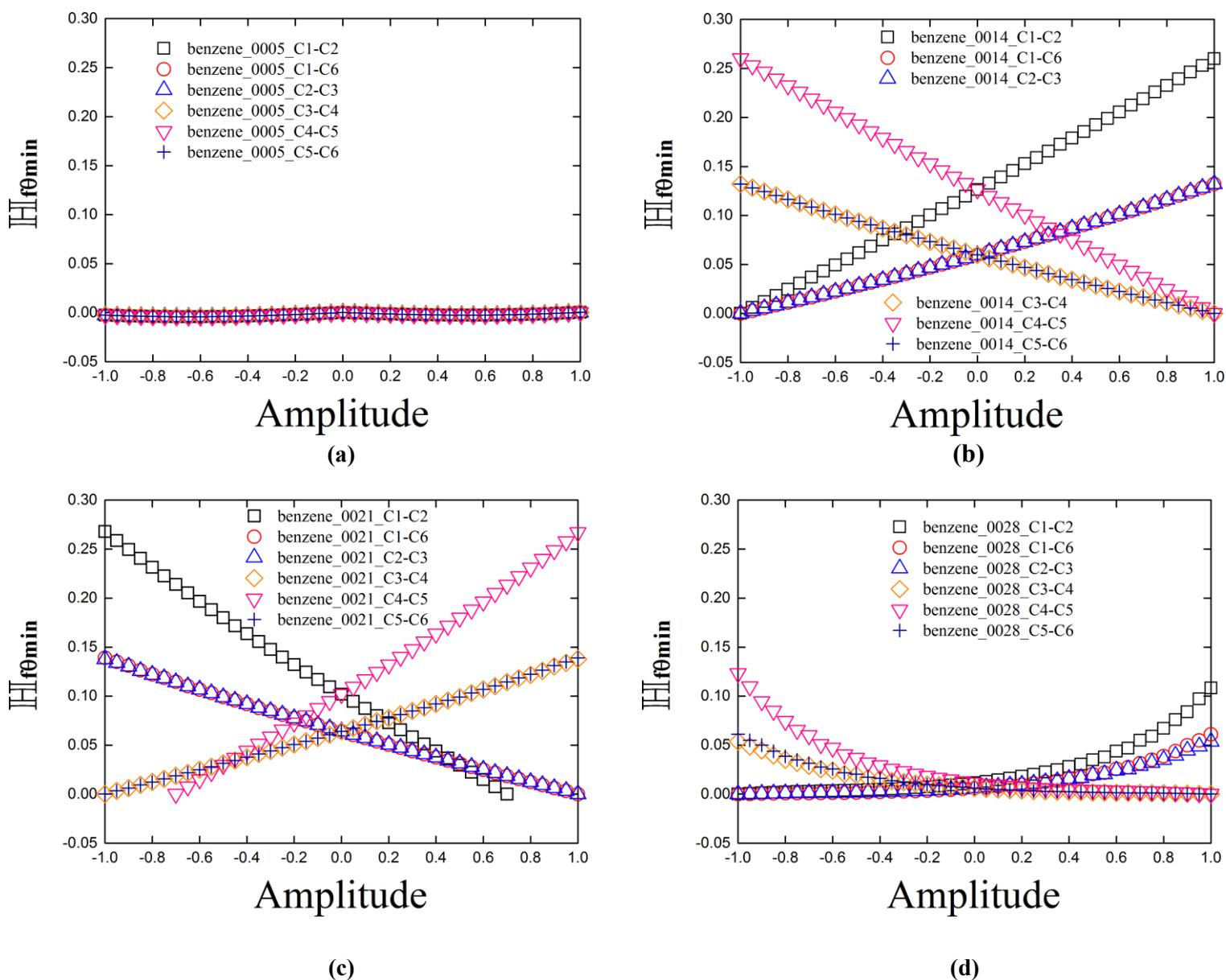

**Figure 5**. The variation of the $\mathbb{H}_{f\theta min} = (\mathbb{H} - \mathbb{H}_{\theta min})/\mathbb{H}_{\theta min}$ of the C-C BCPs with amplitude for the four IR active normal modes of benzene are shown in the left and right hand panels of sub-figures **(a)-(d)** respectively, the corresponding values for $\mathbb{H}^*_{f\theta min} = (\mathbb{H}^* - \mathbb{H}^*_{\theta min})/\mathbb{H}^*_{\theta min}$ are provided in the **Supplementary Materials S5**.

Further magnification of the features of the IR benzene normal modes can be seen by inspection of the fractional eigenvector-following path length $\mathbb{H}_{f\theta min}$, for both the C-C *BCP*s and C-H *BCP*s, see **Figure 5** and **Figure 6** respectively, note the difference in scale used for $\mathbb{H}_{f\theta min}$ compared with $\mathbb{H}_f$. Differences between $\mathbb{H}_{f\theta min}$ and $\mathbb{H}^*_{f\theta min}$ for mode 0014, mode 0021 and mode 0028 are now also more apparent.

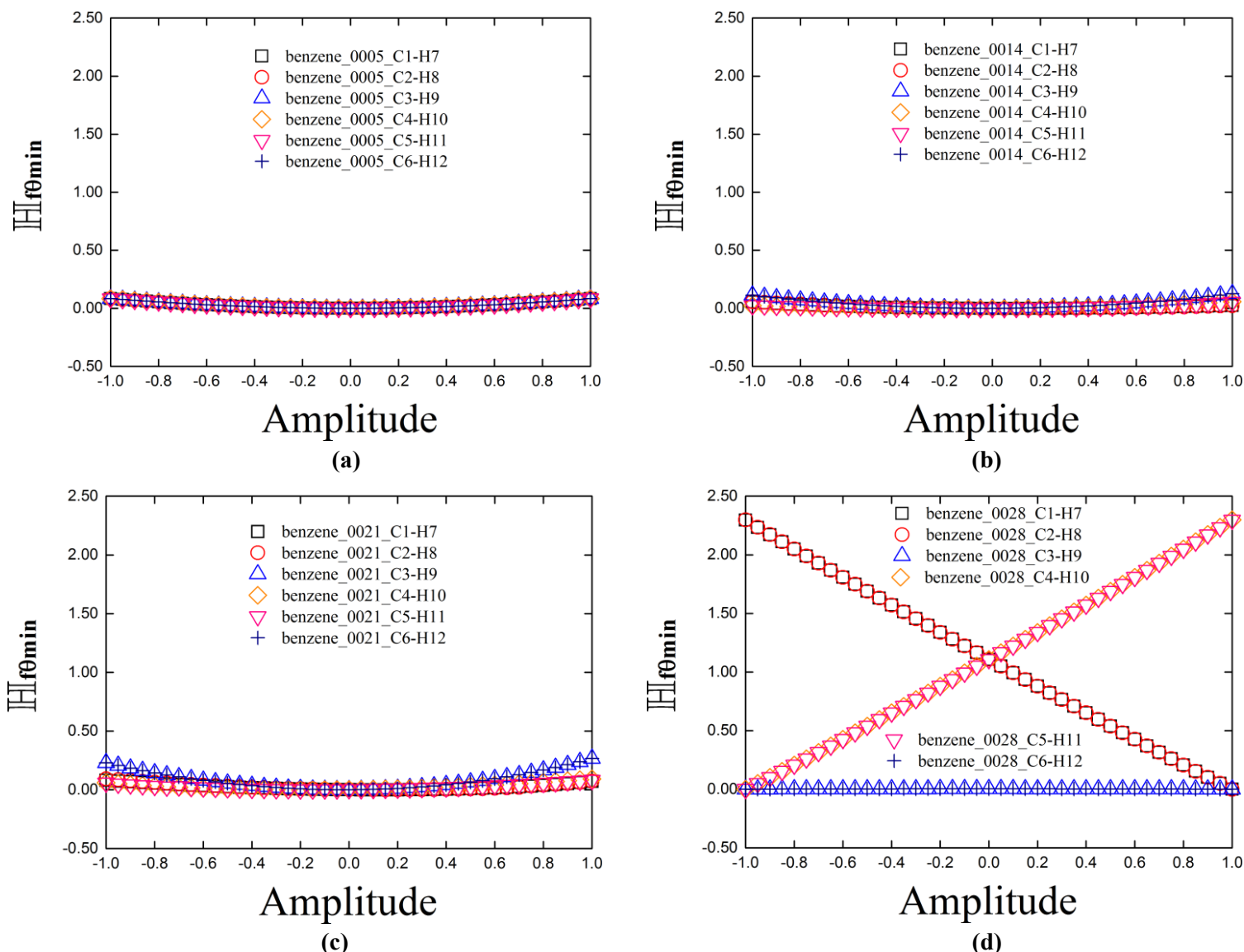

**Figure 6**. The variation of the $\mathbb{H}_{f\theta min} = (\mathbb{H} - \mathbb{H}_{\theta min})/\mathbb{H}_{\theta min}$ of the C-H *BCPs* with amplitude for the four IR active modes of benzene are shown in the left and right hand panels of sub-figures **(a)-(d)** respectively, see the caption of **Figure 5** for further details, the corresponding values for $\mathbb{H}^*_{f\theta min} = (\mathbb{H}^* - \mathbb{H}^*_{\theta min})/\mathbb{H}^*_{\theta min}$ are provided in the **Supplementary Materials S6**.

**Conclusions**

A new QTAIM interpretation of a bond has been established and applied to an analysis of the four IR active modes of benzene. The bond-path framework set $\mathbb{B} = \{p, q, r\}$, with corresponding lengths $\mathbb{H}^*$, $\mathbb{H}$ and BPL respectively, includes the orthogonal triad eigenvector set; **e₁**, **e₂** and **e₃** of the Hessian of $\rho(\mathbf{r})$ in the construction. The bond-path framework set $\mathbb{B}$ was calculated along the entire extent of a bond, i.e. the bond-path as opposed to only using the *BCP*. The variation of the BPL with amplitude was found to capture the stretching/compression motions of the bond-paths associated with the C-C *BCP*s and C-H *BCP*s of the four IR active modes of benzene. The variations of the newly introduced eigenvector-following path lengths and $\mathbb{H}^*$ and $\mathbb{H}$ captured the corresponding torsional bond-path motions as determined by the presence of non-linear variations of $\mathbb{H}^*$ and $\mathbb{H}$ with amplitude. Therefore, the bond-path framework set $\mathbb{B}$ was found to be suitable for capturing the morphology in 3-D of the four benzene IR active normal modes expressed in terms of the $\mathbb{H}^*$, $\mathbb{H}$ and BPL. We found slight differences in the eigenvector-following path lengths and $\mathbb{H}^*$ and $\mathbb{H}$ for the presence of bond-path torsion that occur for non-linear variations of $\mathbb{H}$ with amplitude are due to slight variations from linearity of a bond-path (*r*). The fractional eigenvector-following path $\mathbb{H}_f$ and $\mathbb{H}_{f\theta min}$ and were found to be useful in magnifying the non-linear variations of $\mathbb{H}$ with amplitude whereas the bond-path curvature was not found to usefully magnify the variations of the BPL. This is because of the inability of the BPL and consequently bond-path curvature to detect the presence of torsion of the BPL. This is because $\mathbb{H}^*$ and $\mathbb{H}$ are defined by the precession of the **e₁** and **e₂** eigenvectors about the **e₃** eigenvector, the later defining the bond-path.

We have demonstrated the ability of the eigenvector-following path length $\mathbb{H}$ as part of the bond-path framework set $\mathbb{B}$, to sensitively identify torsional features in the IR active modes of benzene that are not accessible to either conventional analysis using molecular geometries or the BPL. In doing so, we have introduced a new bond-by-bond measure to establish bond-path torsion without the need to use either dihedral angles or any reference directions. Future work could be the analysis of the normal modes of vibration of any molecule or cluster amenable to QTAIM analysis.

**Acknowledgements**

The National Natural Science Foundation of China is gratefully acknowledged, project approval number: 21673071. The One Hundred Talents Foundation of Hunan Province and the aid program for the Science and Technology Innovative Research Team in Higher Educational Institutions of Hunan Province are also gratefully acknowledged for the support of S.J. and S.R.K.

# References

1. M Vollmer, MÃ Klaus-Peter. *Infrared Thermal Imaging: Fundamentals, Research and Applications*. (Wiley-VCH, 2017).

2. P Larkin. *Infrared and Raman Spectroscopy: Principles and Spectral Interpretation*. (Elsevier, 2017).

3. Jenkins, S., Morrison, I. & Ross, D. K. Symmetry classification of the projected vibrational density of states in ice VIII from ab initio methods. *J. Phys. Condens. Matter* **12,** 815 (2000).

4. Hu, M. X. *et al.* The normal modes of vibration of benzene from the trajectories of stress tensor eigenvector projection space. *Chem. Phys. Lett.* **677,** 156–161 (2017).

5. Wilson, E. B. The Normal Modes and Frequencies of Vibration of the Regular Plane Hexagon Model of the Benzene Molecule. *Phys. Rev.* **45,** 706–714 (1934).

6. Margoshes, M. & Fassel, V. A. The infrared spectra of aromatic compounds: I. The out-of-plane C-H bending vibrations in the region 625–900 cm−1. *Spectrochim. Acta* **7,** 14–24 (1955).

7. Satink, R. G., Piest, H., von Helden, G. & Meijer, G. The infrared spectrum of the benzene–Ar cation. *J. Chem. Phys.* **111,** 10750–10753 (1999).

8. Cole, A. R. H. & Michell, A. J. The dipole moment of the C☐H bond in benzene derivatives from infrared intensities. *Spectrochim. Acta* **20,** 739–746 (1964).

9. W Thomson. *Theory of Vibration with Applications*. (Nelson Thornes Ltd, 2018).

10. G Varsanyi. *Vibrational Spectra of Benzene Derivatives*. (Elsevier, 2012).

11. Bader, R. F. W. *Atoms in Molecules: A Quantum Theory*. (Oxford University Press, USA, 1994).

12. Xu, T. *et al.* A stress tensor eigenvector projection space for the (H2O)5 potential energy surface. *Chem. Phys. Lett.* **667,** 25–31 (2017).

13. Guo, H. *et al.* Distinguishing and quantifying the torquoselectivity in competitive ring-opening reactions using the stress tensor and QTAIM. *J. Comput. Chem.* **37,** 2722–2733 (2016).

14. Bell, R. P., Thompson, H. W. & Vago, E. E. Intensities of Vibration Bands. I. Bending Vibrations of Benzene Derivatives and the Dipole of the C-H Link. *Proc. R. Soc. Lond. Ser. Math. Phys. Sci.* **192,**

498–507 (1948).

15. Mulliken, R. S. *Mulliken, R. S. Spectroscopy, Molecular Orbitals, and Chemical Bonding. Science 157, 13–24*. (1967).

16. Alcolea Palafox, M. Scaling factors for the prediction of vibrational spectra. I. Benzene molecule. *Int. J. Quantum Chem.* **77,** 661–684 (2000).

17. Goeppert-Mayer, M. & Sklar, A. L. Calculations of the Lower Excited Levels of Benzene. *J. Chem. Phys.* **6,** 645–652 (1938).

18. Handy, N. C., Murray, C. W. & Amos, R. D. Study of methane, acetylene, ethene, and benzene using Kohn-Sham theory. *J. Phys. Chem.* **97,** 4392–4396 (1993).

19. Shafiee, G. H., Sadjadi, S. A., Najafpour, J., & Shafice, H. The Correlation between Molecular Graph Properties and Vibrational Frequencies. *J Phys Theor Chem* **6(2),** 1–6 (2009).

20. Lord, R. C. & Andrews, D. H. Entropy and the Symmetry of the Benzene Molecule. *J. Phys. Chem.* **41,** 149–158 (1937).

21. Ayers, P. W. & Jenkins, S. An electron-preceding perspective on the deformation of materials. *J. Chem. Phys.* **130,** 154104 (2009).

22. Jenkins, S., Kirk, S. R., Côté, A. S., Ross, D. K. & Morrison, I. Dependence of the normal modes on the electronic structure of various phases of ice as calculated by ab initio methods. *Can. J. Phys.* **81,** 225–231 (2003).

23. Bone, R. G. A. & Bader, R. F. W. Identifying and Analyzing Intermolecular Bonding Interactions in van der Waals Molecules. *J. Phys. Chem.* **100,** 10892–10911 (1996).

24. Bader, R. F. W. A Bond Path: A Universal Indicator of Bonded Interactions. *J. Phys. Chem. A* **102,** 7314–7323 (1998).

25. Bader, R. F. W. Bond Paths Are Not Chemical Bonds. *J. Phys. Chem. A* **113,** 10391–10396 (2009).

26. Jenkins, S. & Heggie, M. I. Quantitative analysis of bonding in 90° partial dislocation in diamond. *J.*

*Phys. Condens. Matter* **12,** 10325–10333 (2000).

27. Jenkins, S. & Morrison, I. The chemical character of the intermolecular bonds of seven phases of ice as revealed by ab initio calculation of electron densities. *Chem. Phys. Lett.* **317,** 97–102 (2000).

28. Frisch, M. J. W. T., H. B. Schlegel, G. E. Scuseria, M. A. Robb, J. R. Cheeseman, G. Scalmani, V. Barone, B. Mennucci, G. A. Petersson, H. Nakatsuji, M. Caricato, X. Li, H. P. Hratchian, A. F. Izmaylov, J. Bloino, G. Zheng, J. L. Sonnenberg. *Gaussian 09, Revision E.01*. (Gaussian, Inc., 2009).

29. Keith, T. A. *AIMAll, Revision 17.01.25*. (TK Gristmill Software, 2017).